# Evidence for energy-dependent scattering dominating thermoelectricity in heavy fermion systems

Daiki Goto[a], Kentaro Kuga[a], Kiyohisa Tanaka[b], Tsunehiro Takeuchi[a], and Masaharu Matsunami[a]*

[a]*Toyota Technological Institute, Nagoya 468-8511, Japan*
[b]*UVSOR Facility, Institute for Molecular Science, Okazaki 444-8585, Japan*
**e-mail:* matunami@toyota-ti.ac.jp

## ABSTRACT

In the field of thermoelectric materials and devices, improving energy conversion efficiency remains a long-standing challenge. As a promising approach to address this issue, utilizing energy-dependent electron-scattering beyond the ordinary constant relaxation time approximation (CRTA) has been proposed. However, direct experimental evidence for an energy-dependent scattering reflected in the Seebeck coefficient is still lacking. Here we demonstrate using angle-resolved photoemission spectroscopy that the relaxation time of heavy fermion quasiparticles is highly dependent on the energy near the Fermi level. The observed energy dependence of the relaxation time is due to the coherent Kondo scattering, describing the sign of the Seebeck coefficient reasonably well, which cannot be deduced from CRTA. Our findings provide not only deeper insight into the understanding of thermoelectricity in correlated materials, but also future perspectives on possible orbital-selective engineering of thermoelectric materials.

## MAIN TEXT

Thermoelectric technology based on the Seebeck effect, which is capable of converting temperature differentials directly into electrical energy, can maximize the efficient use of existing energy sources and hence promote the development of a sustainable economy and society. The bottleneck preventing the widespread adoption of this promising technology is its conversion inefficiency. As evaluation criteria for energy conversion efficiency, we use the figure of merit ($ZT$) or power factor (PF):

$$ZT = \frac{\mathrm{PF}}{\kappa} T, \qquad \mathrm{PF} = S^2 \sigma \tag{1}$$

where $\kappa$ is the thermal conductivity, $T$ is the absolute temperature, $S$ is the Seebeck coefficient or thermopower, and $\sigma$ is the electrical conductivity. To enhance $ZT$, which signifies improved energy conversion efficiency, a combination of high PF and low $\kappa$ is required. Regardless of $ZT$, boosting PF itself can be very important in cases of power generation between thermal reservoirs at fixed temperatures[1] or active cooling.[2] To enhance PF, it is necessary to amplify $S$ while keeping $\sigma$ high, which requires optimizing $S$ beyond the ordinary rigid-band model. On the basis of the Mott equation,[3] $S$ of metals or heavily doped degenerate semiconductors is well described as:

$$S = -\frac{\pi^2 k_{\mathrm{B}}^2 T}{3e}\left[\frac{\partial \ln(D(E))}{\partial E} + \frac{\partial \ln(\tau(E))}{\partial E}\right]_{E=E_{\mathrm{F}}} \quad (2)$$

where $k_{\mathrm{B}}$ is Boltzmann constant, $e$ is the unit of charge, and $E_{\mathrm{F}}$ is the Fermi energy. $D(E)$, and $\tau(E)$ represent the density of states (DOS) and relaxation time (lifetime) of electrons, respectively, at a given energy $E$. Here the energy dependence of the group velocity $v_G(E)$ is ignored, as described later. In many cases, the second term regarding $\tau(E)$ is also ignored in describing $S$ since $\tau$ is expected to be independent of the energy near $E_{\mathrm{F}}$. Indeed, such a constant relaxation time approximation (CRTA) has been successfully used to predict and interpret the transport properties of well-known thermoelectric materials.[4-7] In several materials exhibiting high PF, on the other hand, a non-trivial contribution of $\tau(E)$ to $S$ has been anticipated due to electron-phonon scattering in CoSi,[8] layered perovskite $Sr_2Nb_2O_7$,[9] half Heusler alloys[10] and germanene;[11] energy-dependent hopping conduction in $CuFeS_2$;[12] and interband scattering in NiAu alloys.[13] In these studies, unfortunately, the impact of $\tau(E)$ is considered not on the basis of direct evidence but rather just to explain thermoelectric properties consistently or by theoretical predictions. While $\tau(E)$ engineering might provide an alternative avenue for enhancing thermoelectric performance, direct experimental evidence for $\tau(E)$ is still lacking. As a related problem, the benefits in PF due to $\tau(E)$ can be achieved as well as for $D(E)$ when a sharply varying $D(E)$ exists asymmetrically with respect to $E_{\mathrm{F}}$. This is due to the close correlation between $\tau(E)$ and $D(E)$, causing confusion in determining the dominant factor in $S$. Such a situation is typically seen in heavy fermion systems, as follows.

In heavy fermion systems, the effective mass of conduction ($c$-) electrons can be highly enhanced through hybridization with $4f$ electrons localized near $E_{\mathrm{F}}$. Such a "heavy electron" feature provides a large $S$ coexisting with good metallicity, resulting in a large PF.[14] Heavy-fermion systems have therefore been promising candidates for next-generation thermoelectric materials.[15] As shown in Fig. 1a, indeed, typical heavy fermion systems have a relatively large $S(T)$ despite their metallic nature with an electrical resistivity less than 230 μΩcm.[16-21] Generally, heavy fermion Ce-based and Yb-based compounds are of p-type and n-type, respectively, over a wide temperature range.[22-23] These characteristics are related to the energy position of the Kondo resonance peak in DOS just above and below $E_F$, as shown in Fig. 1b, reflecting their electron-hole symmetry: one electron for $Ce^{3+}$ ($f^1$) and one hole for $Yb^{3+}$ ($f^{13}$) states. However, the sign of $S(T)$ cannot be described in terms of $D(E)$. Considering the slope of the Kondo resonance peak at $E_F$ and applying it in Eq. 2 while ignoring the second term, we obtain $S(T)$ with the opposite sign to the measured one. In such a case, $\tau(E)$ near $E_F$ should be taken into account, for example as shown in Fig. 1c. In the heavy fermion cases, the Kondo scattering of *c*-electrons by 4*f* electrons possibly plays a dominant role in $\tau(E)$. Such a scenario has been described theoretically[24] and sometimes considered in analyzing thermoelectric properties,[25] but no experimental evidence for $\tau(E)$ has been reported so far. In estimating $\tau(E)$, a simple inverse DOS model, more precisely, a model using the inverse partial DOS of *d* or *f* electrons as scatters, has been supposed in broad classes of materials not just for heavy fermion systems, but it is limited to just a qualitative comparison,[8] suggesting the importance of direct observation of $\tau(E)$.

In this work, we experimentally investigated $\tau(E)$ of quasiparticles in heavy fermion systems by using angle-resolved photoemission spectroscopy (ARPES). The linewidth of the band dispersions probed by ARPES is directly related to the quasiparticle lifetime, hence $\tau(E)$ can be evaluated. Owing to the Fermi-Dirac distribution, we can observe the thermally occupied part up to energies of a few $k_BT$ above $E_F$, which affects the thermoelectric properties. As a sample, we selected a prototypical heavy fermion system, $YbCu_2Si_2$, with a layered structure as shown in Fig. 2a, possessing a relatively large $|S|$, as shown in Fig. 1a. The obtained $\tau(E)$ describes the sign of $S$ reasonably well, which cannot be deduced from CRTA. Such an experimental determination of $\tau(E)$ is applicable to the broad classes of thermoelectric materials.

Single crystals of $YbCu_2Si_2$ were prepared by the Sn-flux method. The starting elements were Yb (3N: 99.9% pure) Cu (4N), Si (5N), and Sn (3N). These elements, with an off-stoichiometric ratio of Yb : Cu : Si : Sn = 1 : 15 : 2 : 50 were inserted in an alumina crucible and sealed in a quartz tube. The quartz tube was heated to 1323 K, held at this temperature for 2 days, and cooled at 2 K/h to 873 K, over about 12 days. The excess flux was removed in an external centrifuge. $S(T)$ was measured by the steady-state method using a commercial physical property measuring system.

Synchrotron ARPES measurements were carried out at the undulator beamline BL5U of the UVSOR facility in the Institute for Molecular Science, using photon energies of 40-121 eV and a hemispherical electron analyzer, MBS-A1. The total energy resolution was set to 21 meV for a photon energy of 40 eV and 92 meV for 121 eV. The clean (001) surfaces of single crystals were obtained by *in-situ* cleaving under ultra-high vacuum at a measurement temperature of 8.5 K. $E_F$ was determined by measuring the photoemission spectra of a polycrystalline gold film in electrical contact with the samples.

In the present study, the ARPES image is measured along the directions from Γ to X or from Z to X in the neighboring Brillouin zone, as shown in Fig. 2b, corresponding to the $\bar{\Gamma}$ - $\bar{X}$ direction in the surface Brillouin zone. Figure 2c shows the ARPES image over the wide energy range of $YbCu_2Si_2$ measured with a photon energy of 121 eV. In addition to the free electron-like parabolic dispersions of *c*-bands consisting of Yb 5*d*, Cu 3*d*, and Si 3*p* electrons centered at X point,[20] the two dispersion-less Yb 4*f* bands are clearly observed near $E_F$ ($4f_{7/2}$) and at 1.4 eV ($4f_{5/2}$) split by spin-orbit interactions, together with the respective surface components at 0.6 and 1.9 eV. Hereafter, we focus on the bands of $4f_{7/2}$ and *c*-electrons near $E_F$ and their hybridization, mainly contributing to the thermoelectric properties. The *c*-band dispersions estimated from the peak maxima of momentum distribution curves (MDCs) are almost independent of the photon energy, as shown in Fig. 2d, suggesting a quasi-two-dimensionality of this band structure as discussed later.

Figure 3a shows the ARPES image measured with a photon energy of 40 eV, which is of higher energy and momentum resolution. To better visualize the band dispersions and $\tau(E)$, we performed detailed analyses as follows. The band dispersions are simply represented by the peak maxima estimated from the Lorentzian fitting for MDCs and energy distribution curves (EDCs). The scattering rate, $\Gamma$, as an inverse $\tau$, is closely related to the self-energy, $\Sigma$, which is contained in the single-particle spectral function and accounts for the many-body effects. In ARPES, an imaginary part of the self-energy, $\mathrm{Im}\Sigma$, is evaluated not just from the peak width of EDCs, $\Delta\omega$, but also from that of MDCs, $\Delta k$, through the relation:[26]

$$2\mathrm{Im}\Sigma \approx \Gamma = \frac{\hbar}{\tau} = \hbar v_G \Delta k \tag{3}$$

Therefore, $\Gamma$ (and also $\tau$) is determined by multiplying $\Delta k$ with the group velocity $v_G$ at a given energy. Figure 3b shows the typical results of Lorentzian fitting for the selected

MDCs. The energy-dependent variation of the peak width and the peak position is clearly confirmed even at energy slightly above $E_F$. The dispersions obtained from the peak maxima are superimposed on the ARPES image in Figs. 3c and 3d. To simplify the analysis, we approximated the band dispersions with a linear relation ignoring a deviation around the intersection, as indicated by the solid line in Fig. 3c. In this case, the energy dependence of $v_G$ near $E_F$ can be ignored, and hence the term of $v_G$ was not involved in Eq. 2.

The highly dispersive bands estimated from MDCs in Fig. 3c are due mainly to *c*-electrons, and less dispersive band estimated from EDCs to 4*f* electrons. These bands, particularly the *c*-bands, gradually bend toward the intersection by mixing with each other, providing evidence of *c*-*f* hybridization. The *c*-*f* hybridized bands can be typically seen in heavy fermion systems[27-31] and are well explained in terms of a hybridization model based on the periodic Anderson model,[32] as schematically drawn in Fig. 3e. Through *c*-*f* hybridization, the Yb 4*f* electrons affect the Fermi surface, giving rise to valence fluctuations in Yb ions.[33] Such hybridization also affects the thermoelectric properties by modulating the *c*-band dispersions near $E_F$.

As we expected, $\Delta k$ exhibits a strong energy dependence as shown in Fig. 4a. A significant variation in $\Delta k(E)$ is seen on both sides of the Kondo resonance peak in the angle-integrated spectra qualitatively representing DOS in Fig. 4b, indicating that the Kondo effect plays a key role in this scattering. The ground state of heavy fermion systems is a Fermi liquid with the local moment screened by *c*-electrons. At a temperature well below the Kondo temperature as here, the dominant interaction is the coherent scattering of *c*-electrons on the 4*f* ``lattice'', which is distinguished from the single-site Kondo effect at a higher temperature. The observed energy-dependent scattering, for which $\Delta k$ is strongly suppressed toward $E_F$ along the slope of the Kondo resonance peak, is due to the coherent Kondo scattering. Such a feature also provides the strong increase in $\tau(E)$ toward the unoccupied states above $E_F$ as shown in Fig. 4c. The resulting $\tau(E)$ contributes negatively to the sign of $S(T)$ through the Eq. 2 in contrast to $D(E)$, as expected in Figs. 1b and 1c. As a consequence, we certainly confirm that the dominant factor in $S(T)$ of heavy fermion systems is due not to the first term regarding $D(E)$ in Eq. 2, but to the second term regarding $\tau(E)$. On the other hand, as shown in Fig. 4d, the inverse $D(E)$ indicates a similar behavior to $\tau(E)$, ignoring minor differences. Therefore, the inverse DOS model for describing $\tau(E)$ can be applicable in a qualitative sense but should be experimentally verified.

We discuss the possibility of reproducing $S(T)$ on the basis of our ARPES data and analyses. In this regard, the following issues can be anticipated: (i) impact of the final-state broadening on $\Gamma$, (ii) identification of the bulk/surface electronic structure and the need to observe all bands in the Brillouin zone. For three-dimensional materials, the final-state broadening on $\Gamma$ is unavoidable in photoemission experiments,[34] particularly with use of low photon energy, as here. Compared with $\tau$ calculated from the Drude model and transport data as described in the supplementary material, indeed, $\tau(E = E_F)$ in Fig. 4b is three orders of magnitude shorter, suggesting an impact of the final-state broadening of $\Gamma$. Even though the absolute value of $\tau$ requires further consideration, the energy dependence itself is still reliable, because the final-state broadening is nearly independent of the energy.

The quasi-two-dimensionality of the *c*-*f* hybridized bands observed here (Fig. 2d) can be interpreted in both terms of surface-derived or bulk two-dimensional features based on the layered structure.[28] In any case, even though obtaining the absolute value of $S$ is challenging, it is highly evident that $S$ is certainly dominated by $\tau(E)$ when the band structure peculiar to heavy fermion Yb compounds,[27-31] which can be described by the *c*-*f* hybridization model, is realized as here. According to DFT-LDA calculations, $YbCu_2Si_2$ has at least two Fermi surfaces, in which one is due to quasi two-dimensional (33th) electron band centered at X point focused in the present study and the other is due to three-dimensional (32th) hole band around Z point.[20] By analyzing just the former electron band, which is clearly observable, we can provide an important insight into the sign reversal problem of $S$ regarding the Kondo resonance peak. To completely quantify $S(T)$ in the future, in addition to the temperature dependent ARPES measurement, the analyses carried out in this work should be applied to all bands (or orbitals) contributing to the Fermi surfaces, which could lead to possible orbital-selective engineering in thermoelectric materials.

In summary, we investigated the role of energy-dependent scattering on the thermoelectric properties in heavy fermion systems by means of ARPES. From the linewidth of the band dispersions, particularly in MDCs, the energy-dependent scattering and $\tau(E)$ of quasiparticles in $YbCu_2Si_2$ are evaluated. A significant variation in $\tau(E)$ is seen around the energy of the Kondo resonance peak, reflecting Kondo scattering. This energy dependence describes the sign of $S$ reasonably well, which cannot be deduced from CRTA. This method is applicable to various materials possessing high PF, providing an alternative strategy for further enhancing the performance.

See the supplementary material for the comparison with DFT calculations, the energy dependence of relaxation time estimated from higher-photon energy ARPES data, and the relaxation time estimated from Drude model.

## ACKNOWLEDGEMENT

We would like to thank K. Kandai, M. Okada and F. Ogawa for collaboration in the early stage of this work, S. Uchida for help in the ARPES measurement, and Prof. H. Harima for providing us with the results of the band calculations. This work was supported by JSPS KAKENHI Grant No. 23K23051. The synchrotron ARPES was performed by the Use-of-UVSOR Facility Program of the Institute for Molecular Science (2024).

## DATA AVAILABILITY

The data that support the findings of this study are available from the corresponding author upon reasonable request.

## FIGURES

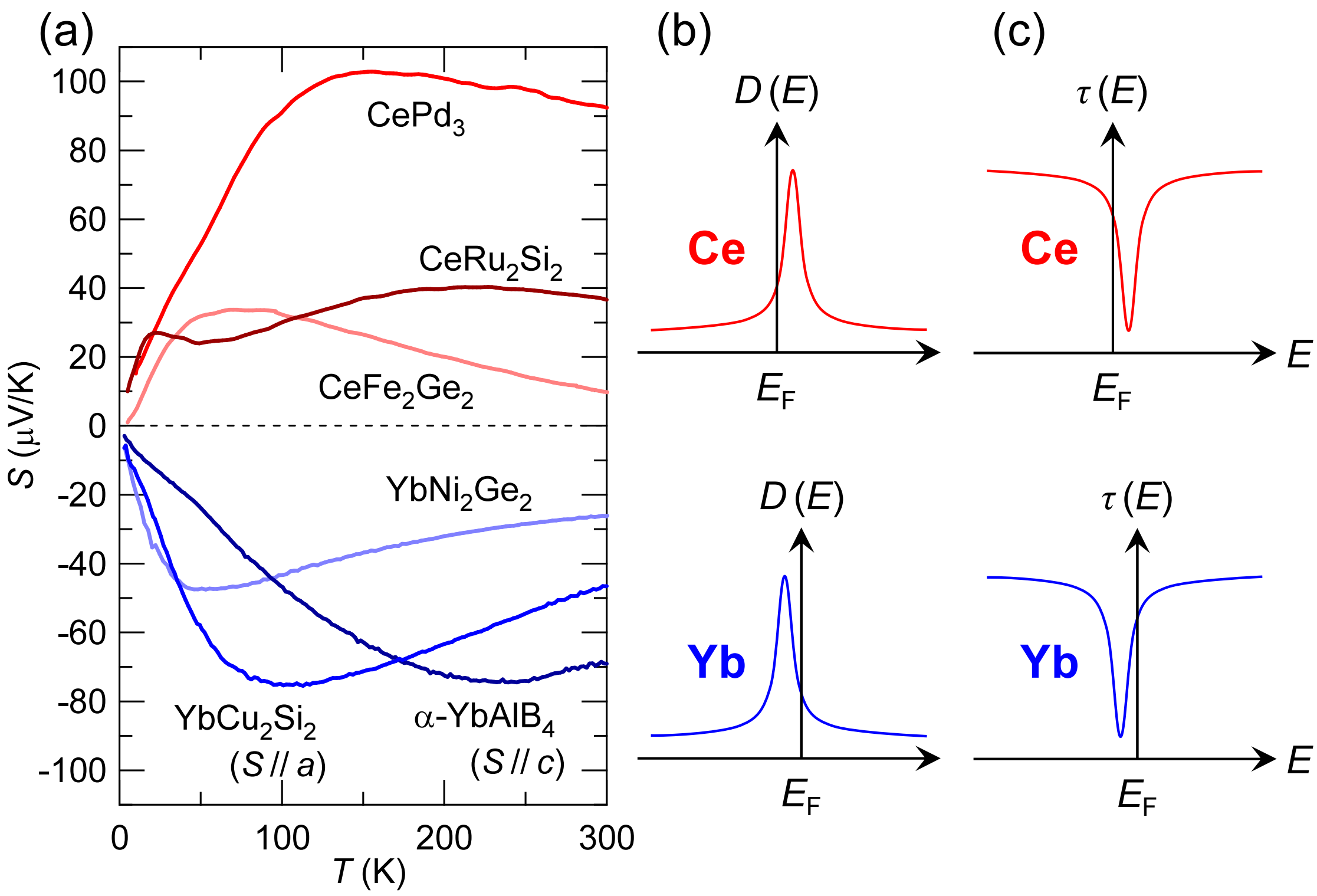

**Fig. 1.** (a) Seebeck coefficient $S(T)$ of typical heavy fermion Ce- and Yb-based compounds: polycrystalline $CePd_3$, $CeRu_2Si_2$, $CeFe_2Ge_2$, and $YbNi_2Ge_2$, which were prepared by the arc melting method, and single crystalline α-$YbAlB_4$[21] and $YbCu_2Si_2$. (b) Schematic DOS $D(E)$ in the vicinity of $E_F$ for heavy fermion Ce- and Yb-based compounds. The Kondo resonance peak lies just above $E_F$ in Ce-based compounds and just below $E_F$ in Yb-based compounds. (c) Schematic energy dependence of relaxation time $\tau(E)$ expected in the inverse DOS model.

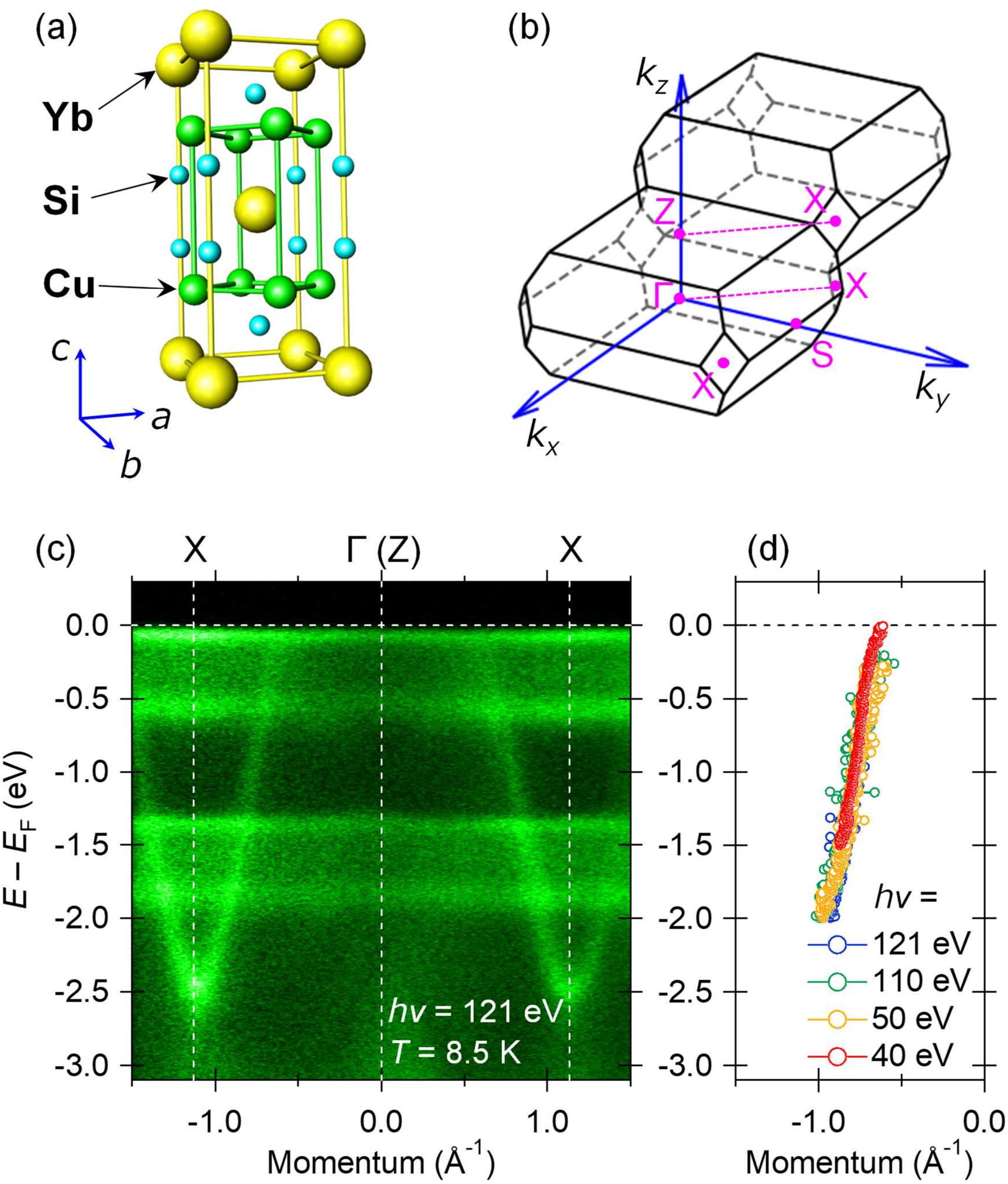

**Fig. 2.** (a) Tetragonal $ThCr_2Si_2$-type crystal structure. (b) Corresponding Brillouin zone. The present ARPES data are focused along the Γ-X or Z-X direction. (c) ARPES intensity and energy-momentum distribution map measured with a photon energy of 121 eV. (d) Band dispersions measured with photon energies of 40, 50, 110, and 120 eV. The data points are the maxima in MDCs at each energy.

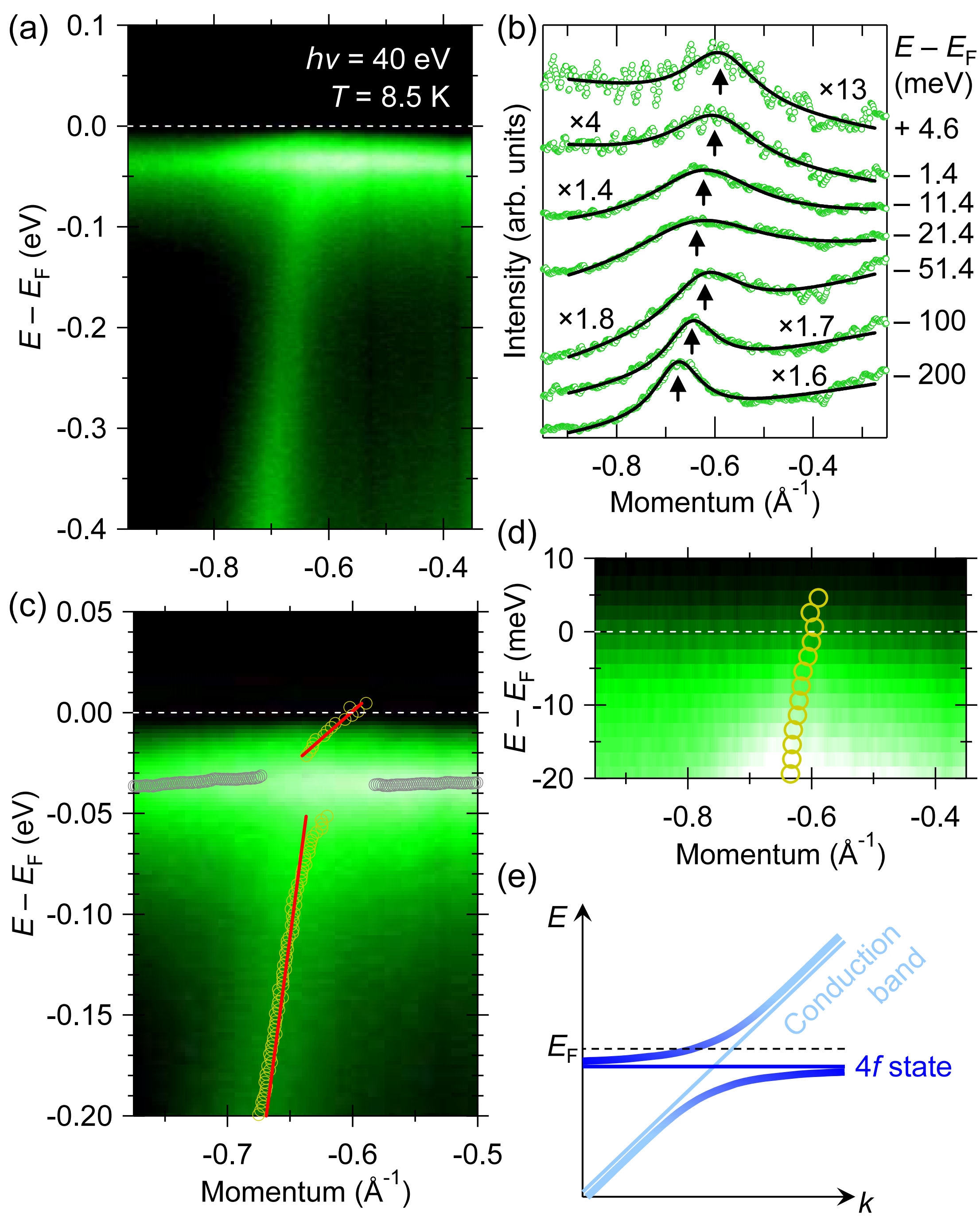

**Fig. 3.** (a) ARPES image of $YbCu_2Si_2$ measured with a photon energy of 40 eV. (b) Examples of Lorentzian fitting for selected MDCs. (c) Magnified view of (a). The overlaid points are the maxima in MDCs (yellow) and EDCs (gray) evaluated by fitting. The red lines show the linearly approximated dispersions. (d) Another magnified view of (a). The contrast is changed to clarify the band toward the unoccupied states just above $E_F$. (e) Schematic band dispersions of the *c*-*f* hybridization model.

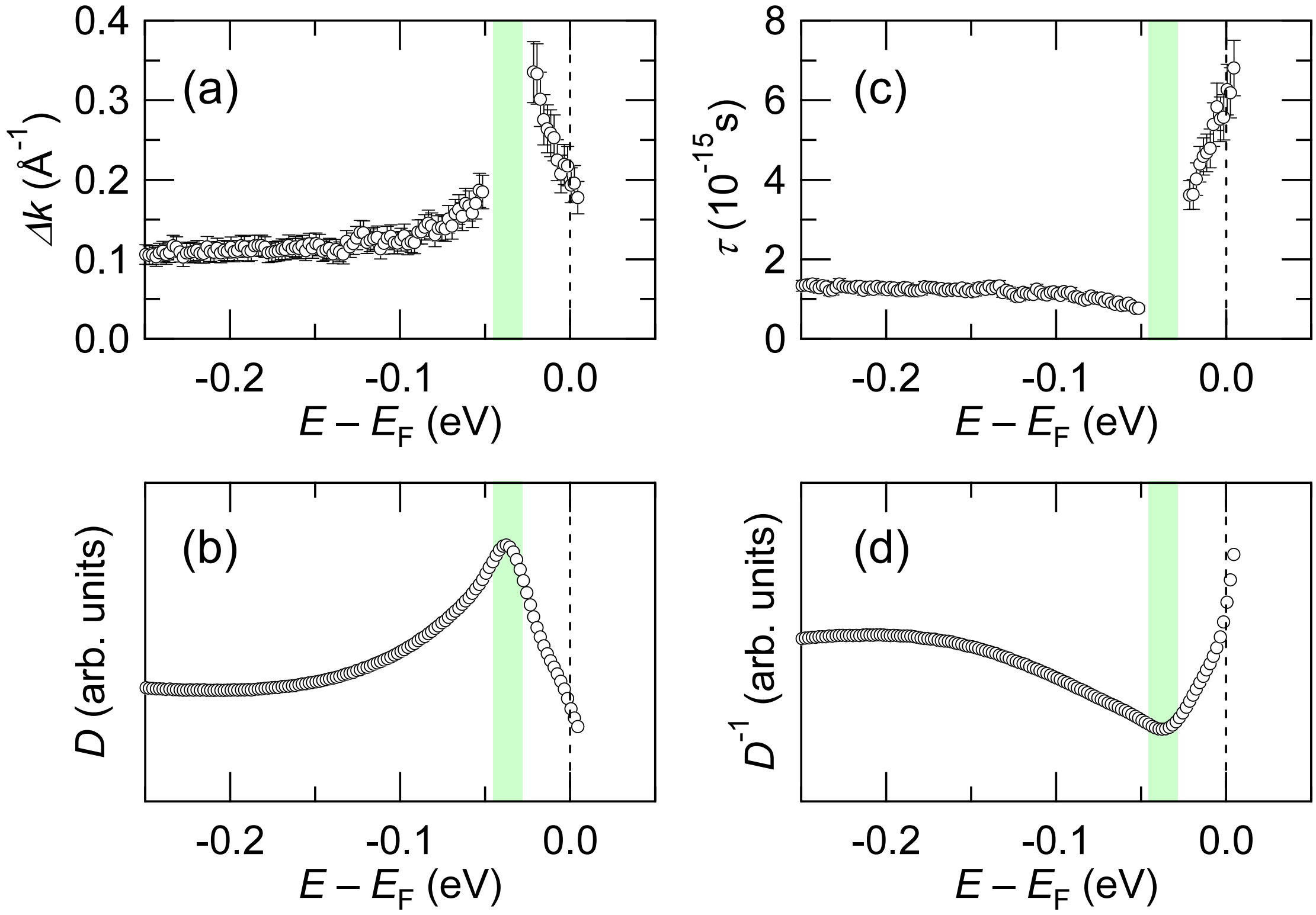

**Fig. 4.** (a) $\Delta k$ (full width at half maximum: FWHM) of MDCs in Fig. 3. (b) Spectral DOS obtained by dividing the momentum integrated ARPES spectra in Fig. 3a by resolution convoluted Fermi-Dirac distribution function. (c) $\tau(E)$ calculated using $\Delta k$ and $v_G$. (d) Inverse DOS calculated from (c). The error bars are from the measurement uncertainties. Green shading in all panels represents the position of the Kondo resonance peak in (b).